\documentclass[a4paper,UKenglish,cleveref, autoref, thm-restate]{oasics-v2021}

\hypersetup{
    colorlinks=true,
    linkcolor=blue!60!black,
    urlcolor=blue!60!black,
    citecolor=blue!60!black
}

\lstdefinestyle{code}{
    basicstyle=\ttfamily\small,
    breaklines=true,
    frame=single,
    columns=fullflexible,
    keepspaces=true,
    showstringspaces=false,
    backgroundcolor=\color{gray!6},
    rulecolor=\color{gray!40},
    keywordstyle=\color{blue!70!black},
    commentstyle=\color{green!40!black},
    stringstyle=\color{red!60!black}
}

\lstdefinestyle{textdiagram}{
    basicstyle=\ttfamily\small,
    breaklines=true,
    frame=single,
    columns=fullflexible,
    keepspaces=true,
    showstringspaces=false,
    backgroundcolor=\color{gray!6},
    rulecolor=\color{gray!40}
}

\usepackage[T1]{fontenc}
\usepackage[utf8]{inputenc}
\usepackage{lmodern}
\usepackage{microtype}
\usepackage{hyperref}
\usepackage{xcolor}
\usepackage{booktabs}
\usepackage{graphicx}
\usepackage{array}
\usepackage{amsmath,amssymb}
\usepackage{stmaryrd} 
\usepackage{listings}
\usepackage{tikz}
\usetikzlibrary{arrows.meta,positioning,shapes,fit,calc}

\usepackage{todonotes}

\hypersetup{
  colorlinks=true,
  linkcolor=blue!60!black,
  citecolor=blue!60!black,
  urlcolor=blue!60!black
}

\lstdefinestyle{code}{
  basicstyle=\ttfamily\small,
  columns=fullflexible,
  breaklines=true,
  frame=single,
  framerule=0.4pt,
  rulecolor=\color{black!25},
  backgroundcolor=\color{black!2},
  keywordstyle=\color{blue!60!black},
  commentstyle=\color{green!40!black},
  stringstyle=\color{orange!50!black},
  showstringspaces=false
}
\newcommand{\cfg}[1]{\ensuremath{\langle #1 \rangle}}

\title{Towards Model-based Run-time Cybersecurity:
\texorpdfstring{\\}{ }
On Control-Flow Anomaly Detection, Attack Identification, and Hardware Monitoring}

\titlerunning{Towards Model-based Run-time Cybersecurity}

\author{Martin Sachenbacher\footnote{corresponding author}}{Faculty of Computer Science and Mathematics, OTH Regensburg, Galgenbergstraße 32, 93053 Regensburg, Germany}{martin.sachenbacher@oth-regensburg.de}{https://orcid.org/0000-0002-5418-1885}{}

\author{Martin Leucker}{Institute for Software Engineering and Programming Languages, Universität zu Lübeck, Ratzeburger Allee 160, 23562 Lübeck, Germany}{leucker@isp.uni-luebeck.de}{https://orcid.org/0000-0002-3696-9222}{}

\author{Alexander Weiss}{Accemic Technologies GmbH, Franz-Huber-Straße 39, 83088 Kiefersfelden, Germany}{aweiss@accemic.com}{https://orcid.org/0000-0003-1029-1297}{}

\author{Aliyu Tanko Ali}{Institute for Software Engineering and Programming Languages, Universität zu Lübeck, Ratzeburger Allee 160, 23562 Lübeck, Germany}{aliyu.ali@isp.uni-luebeck.de}{https://orcid.org/0000-0003-3841-2054}{}

\authorrunning{M. Sachenbacher, M. Leucker, A. Weiss, and A. Ali} 

\Copyright{Martin Sachenbacher, Martin Leucker, Alexander Weiss, and Aliyu Tanko Ali} 



\category{} 

\relatedversion{} 

\supplement{}

\nolinenumbers 

\begin{document}

\maketitle

\begin{abstract}
Methods to increase the resilience of systems to cyber-attacks become increasingly important.
Control-flow monitoring provides a principled
basis to ensure integrity and detect possible anomalies at run-time. Once anomalies have been detected, so-called attack trees can be used to identify possible types of attacks.
However, this approach is vulnerable to camouflage, by which attackers try to evade detection (and correct identification) by deliberately manipulating also the system's observed control flow.
In this paper, we outline a model-based approach that provides more robust intrusion detection and attack identification through an architecture that combines software- with hardware-based monitoring.
In this approach, software-level observation indicates suspicious activities, while
hardware-level monitoring checks them separately in more detail, making it much harder for attacks to camouflage themselves and go undetected.
We illustrate the approach with an authentication-service example that captures a realistic failure mode: a software-level observer sees an
anomalous but apparently harmless control-flow deviation, maps it to a benign
root cause in an attack tree, but misses the true intrusion. A second,
independent hardware control-flow monitor observes the actual transition sequence
and thereby changes the attack-tree diagnosis from a low-severity configuration
or maintenance issue to a high-confidence code-injection or control-flow hijack.
In this scenario, the proposed combination of control-flow anomaly detection, attack-tree based intrusion identification, and hardware-based monitoring can improve not only anomaly detection, but also the diagnostic precision of attack-tree-based cyber-attack identification.
\end{abstract}

\newpage

\section{Introduction: Cybersecurity and Run-time Monitoring} 

Modern computing systems are increasingly exposed to cyber-attacks that target
not only data confidentiality and availability, but also the integrity of the
software execution itself. As software systems become more interconnected,
adaptive, and safety-critical, it is no longer sufficient to protect only their
external interfaces. Instead, resilience must also be provided at run time, when
an adversary may already have gained a foothold inside the system and attempts
to influence its execution. This motivates monitoring techniques that observe
the actual behavior of a system during operation and detect deviations from
expected behavior.

An important class of run-time monitoring techniques is based on
control-flow observation. The control flow of a program describes the sequence of
executed basic blocks, branches, calls, returns, and other control-transfer
events. Many attacks necessarily affect this sequence. For example, code
injection, return-oriented programming, unauthorized privilege escalation,
malicious use of maintenance functionality, or bypasses of authentication logic
may all manifest as control-flow deviations. If a suitable model of legitimate
control flow is available, then deviations from this model can be treated as
anomalies and used as indicators of a possible intrusion. This is the basic
idea behind \emph{control-flow integrity} (CFI), introduced by
Abadi et al.~\cite{Abadi2005}, and behind a sizeable family of hardware-assisted
CFI mechanisms that has emerged since, surveyed in~\cite{declercq2017survey}.

However, for an operator or an automated response/defense mechanism, it is also necessary to interpret the anomaly, i.e. to determine whether the deviation was caused by a benign configuration issue, a rare but legitimate maintenance path, a software defect, or an active attack. One known method
for such interpretation is the use of so-called attack trees \cite{schneier1999attacktrees}. Attack trees decompose attack goals into several subgoals and concrete attack steps.
When combined with observed control-flow evidence, they can support a form of
diagnosis or explanation: an anomalous transition or execution path can be mapped to possible
root problems, and these can in turn be associated with different
attack classes and severities.

This combination of control-flow anomaly detection and attack-tree based
diagnosis is promising, but also vulnerable to a fundamental limitation: apart from possible incorrectness of the model, the diagnosis depends on the quality and correctness of the observation. If the observed control flow is incomplete, manipulated, or misleading, then the attack tree may lead to an incorrect conclusion.
Specifically, attackers may actively attempt
to camouflage an intrusion by shaping either the actual execution or the
monitoring evidence in such a way that the resulting anomaly appears harmless.
This leads to the complication -- conceptually and practically distinct from 'classical' diagnosis of faulty physical systems -- that the observations used for
diagnosis may be \emph{adversarially influenced}.

For example, malicious execution may be hidden inside error-handling,
diagnostic, update, or maintenance paths; alternatively, software-based
instrumentation may be disabled, bypassed, or tampered with so that the
reconstructed control flow omits the decisive malicious transition. In such
cases, the system often may still be able to detect that something unusual happened, but the
subsequent diagnosis will classify the event as a low-severity configuration
problem or benign operational irregularity, rather than as an actual high-severity
intrusion.

In this paper, we aim to address this problem by outlining an architecture for robust, run-time 
intrusion detection and attack identification.
The central idea is to combine software- with hardware-based monitoring: software-level monitoring is used to indicate suspicious activities, while
hardware-level monitoring is used checks them separately in more detail. It monitors the
control flow using a second observation system that is physically independent of the
software system under observation. Thus, rather than relying solely on instrumentation,
logging, or monitoring components that execute inside the potentially compromised
software environment, the proposed approach uses hardware-level trace information about
control transfers. This separation improves the trustworthiness of the
observation channel: an attacker who has compromised the monitored software may
still influence the program's behavior, but it becomes significantly harder to
also manipulate the independent control-flow evidence in a consistent and
undetectable manner.


To illustrate the approach, we describe a concrete example that captures a
realistic failure mode. A software observer detects an anomalous but apparently
harmless control-flow deviation and maps it, via an attack tree, to a benign
maintenance or configuration issue. The true cause, however, is a camouflaged
intrusion in which the attacker has manipulated the software-level observation so
that the decisive malicious transition is not visible. A second, independent
hardware control-flow monitor observes the actual transition sequence. Using
this more truthful trace, the same attack-tree analysis changes its diagnosis
from a low-severity operational explanation to a high-confidence indication of
code injection or control-flow hijacking.

Our work is preliminary as we have implemented parts of this architecture, but not fully integrated them yet. Still, this paper makes three contributions: First, it analyzes the
interplay between run-time control-flow anomaly detection and attack-tree based intrusion
diagnosis. Second, it identifies camouflage of control-flow observations as a
failure mode in which an anomaly may be detected but misclassified. Third, it
outlines a hardware-supported monitoring architecture that provides an
independent source of control-flow evidence and thereby enables more faithful
attack-tree based diagnosis.

\section{Diagnostic Model for Intrusion Detection and Attack Identification}
\subsection{Control-Flow Graphs and Anomalies}

A control flow graph (CFG) \cite{Allen70} represents possible paths through a program or process that might be traversed during its execution. The CFG is as a directed graph where the nodes represent basic blocks, and the directed edges represent control flow paths between them.
Formally, a CFG is given by 
$
G = (V,E,v_e,V_f)
$
where $V$ is a finite set of basic blocks or program locations, $E \subseteq V \times V$ is the set of admissible control-flow transfers, $v_e$ is the entry node, and $V
_f$ is a set of exit nodes. 
An example of a control-flow graph is shown in Figure~\ref{fig:nominal-cfg}.

A CFG can be used to detect anomalous behavior in a system, by comparing observed runtime traces with the execution paths possible in $G$.
Formally, a runtime trace is modeled as an ordered sequence
$
\tau = v_0 v_1 \ldots v_n
$
of basic blocks.

\emph{control-flow integrity} (CFI), introduced by
Abadi et al.~\cite{Abadi2005}, uses CFGs as a model of legitimate control flow and checks observed traces for deviations from this model, treating them as anomalies and indicators of possible intrusions. There exist several established methods for CFI: in its simplest form, it ensures that executions follow only paths allowed by a static control-flow graph. More advanced forms also include  dynamic aspects and so-called shadow stacks \cite{burow18} to protect function returns.
For the purpose of this paper, we focus on simple, static CFGs, such that anomalies correspond to  missing or unexpected blocks in the observed execution.



An observed anomaly only indicates that the observed execution path differs from the expected graph, but does not identify possible causes.
The deviation of the execution path may be due to a software bug, misconfiguration, unusual workload, or malicious activity; in this paper, we focus specifically on the latter aspect.

\subsection{Attack Trees and Identification}
\label{subsec:attack-trees}


In the context of security analysis, attack trees~\cite{schneier1999attacktrees} have been proposed as a structured way to model and reason about possible attacks against a system. A basic attack tree can be formalized as a rooted, directed tree whose root node is an attacker top-level goal (such as unauthorized login), and whose internal nodes represent the logical refinement
and decomposition of attack goals into sub-goals and alternative attack steps, combined by logical operators $\mathsf{AND}$/$\mathsf{OR}$.

Attack graphs~\cite{lallie2020} are a more recent, more general concept that allows for multiple root nodes and sharing of nodes within the tree.
Attack trees may be constructed manually or generated semi-automatically from security analysis tools; we do not deal here with the problem of how to obtain them.

There are different ways how attack trees can be used to in the context of control-flow integrity.
We focus on the simple case where the root(s) are observed control-flow anomalies, to be explained by  possible causes (attack steps).

Different observations (root nodes) of the tree being true are then consistent with different internal nodes of the tree being true (= active), so that it functions as a simple diagnostic model: a set of of active internal nodes represents possible attack steps that explain how the attack causes the observed anomaly.
This diagnostic reasoning with attack trees can be further augmented with a notion of minimality  (minimal set of active nodes), although we do not deal with this in this paper.

Mantel and Probst \cite{Mantel2019} discuss the meaning and purpose of attack trees and note that formal semantics were developed after the original, informal notation.
There are also extensions of attack trees with ordering: SAND attack trees \cite{Jhawar2015} add a sequential $\mathsf{AND}$ operator, meaning that the order of attack steps matters. This is especially relevant to analyzing CFG anomalies, because control-flow traces are inherently sequential.

Attack trees can also be further enriched with attributes such as cost or time stamps \cite{ali2024analysis}, which enable to analyze different attack paths, for example, to determine which attacks have minimal or maximal cost, which require the shortest or longest execution time, or which become infeasible once temporal constraints are taken into account.




\subsection{Intrusion Camouflage}

A key complication in cybersecurity, as opposed to 'classical' fault diagnosis for hardware or software systems,
is that attackers may deliberately hide their actions. That is, rather than merely causing an obvious anomalous control-flow path, an attacker may try to make the malicious path look like legitimate behavior, or may also corrupt the observed data from which the control flow is reconstructed by the defender, in order to disguise the attack.
Thus there are two distinct forms of manipulation:

\begin{description}
    \item[\textnormal{\textit{Manipulating the actual control flow}}:] The attacker shapes the malicious path so that it resembles legitimate behavior, such as logging, error handling, debugging, diagnostics, or maintenance.
    \item[\textnormal{\textit{Manipulating the observed control flow}}:] The attacker tampers with telemetry, tracing hooks, logging, or instrumentation so that the reconstructed CFG does not match the real execution.
\end{description}

The first point is known as control-flow camouflage: adversarial shaping or manipulation of the
executed or observed control-flow trace such that malicious behavior appears
benign, policy-compliant, or diagnostically misleading to a monitor
\cite{goktas2014outofcontrol}.
Specifically, \cite{carlini2015effectivenesscfi} introduce control-flow bending (CFB), where an attacker uses vulnerabilities to obtain powerful malicious behavior while still following paths allowed by a static CFG.
Other common camouflage techniques that can interfere with CFG-based detection include:
dead-code padding; opaque predicates; control-flow flattening; fake error-handling paths;  debug, diagnostic, or maintenance-looking branches; dynamic code loading; reflection or indirect calls; hooking or patching instrumentation; or log or trace tampering.

When using attack trees to interpret control-flow anomalies, the diagnosis is
  only as reliable as the observed control-flow evidence.
Thus if an attacker can
  manipulate or bias the observed control-flow graph, the attack-tree diagnosis
  may converge to an incorrect and/or relatively harmless explanation.


Let $G$ be the expected control-flow graph and $\tau = v_0 v_1 \ldots v_n$ the true control-flow trace executed on the monitored system. In practice, a monitor does not observe $\tau$ directly; instead it receives a derived trace through an observation channel.
We distinguish two such channels: a software-based observer, which relies on
instrumentation, logging, or tracing hooks executing inside the monitored
environment, and a hardware-based observer, which reconstructs control-flow
transfers from processor-level trace information independent of the monitored
software. Let $\hat{\tau}_S$ and $\hat{\tau}_H$ denote the traces produced by
these two channels, respectively, and let
\( \mathcal{O}_S(\tau) = \hat{\tau}_S,  \mathcal{O}_H(\tau) = \hat{\tau}_H \) be the corresponding observation functions.

We remark that more precisely, $\hat{\tau}_H$ denotes the block-level trace reconstructed from the raw hardware trace, not the raw trace stream emitted by the monitor itself. A trace decoder uses the program image, symbol or map information to recover executed branch targets and instruction-address intervals, which are then mapped to the basic blocks used in our model.



Under normal conditions both observations are faithful, but an adversary who has compromised the monitored software may corrupt the software-level channel while leaving the hardware-level channel intact. Camouflage of the software observation is therefore modeled as \( \mathcal{O}_S(\tau) \neq \tau, \) 
whereas the hardware monitor is assumed to provide a substantially more faithful reconstruction: \( \mathcal{O}_H(\tau) \approx \tau. \)


An anomaly may therefore be present in the true trace $\tau$ yet absent in the
software-observed trace $\hat{\tau}_S$, absent in the true trace yet present in
$\hat{\tau}_S$ due to corruption, or present in both but at different locations,
depending on how the attacker has shaped the observation.

Given an observed trace $\hat{\tau}$, feature vectors
of anomalies with respect to $G$
can be defined: 
$
F_G(\hat{\tau}) =
  \{\phi_1(\hat{\tau}),\ldots,\phi_m(\hat{\tau})\}
$
where examples of $\phi_i$ are (see Section \ref{sec:example}) missing blocks or unexpected blocks.
Then an attack-tree diagnosis is defined as a function  $
D_T(F_G(\hat{\tau})) \subseteq \mathcal{H}
$
that returns the set $\mathcal{H}$ of attack-tree internal nodes best explaining the
observed trace features.

%
%
%
%
%
%
%
%
%
%
%
%
%
%
%
%
%








\section{Hardware-based Control-Flow Monitoring}
\label{sec:approach}



Our proposed approach, building on previous work in \cite{LeuckerDX2023}, exploits a capability for non-intrusive system monitoring that runs
separately from the system itself. As a consequence, this can be used in applications where
the system is unaware of, or even averse to, being monitored.

 For the goal of detecting and uncovering of cyber-attacks and intrusions in systems,
 hardware-based monitoring can improve the diagnosis by providing an independent observation channel that is harder for software-level malware to suppress or falsify.
 
The idea is to monitor the actual execution, and whenever the reconstructed execution deviates from the validated model (even in the absence of system faults),
it must be due to changes
that have been inflicted on the system after its deployment (e.g., through intrusion, malware, etc.), or a monitoring/modeling error.

Existing software-only approaches for intrusion detection such as virus scanners suffer
from the problem that they are running on the same system as the potential intruders, and
can therefore be compromised themselves.



\begin{figure*}
    \centering
        \captionsetup{skip=12pt}
\includegraphics[width=0.9\columnwidth]{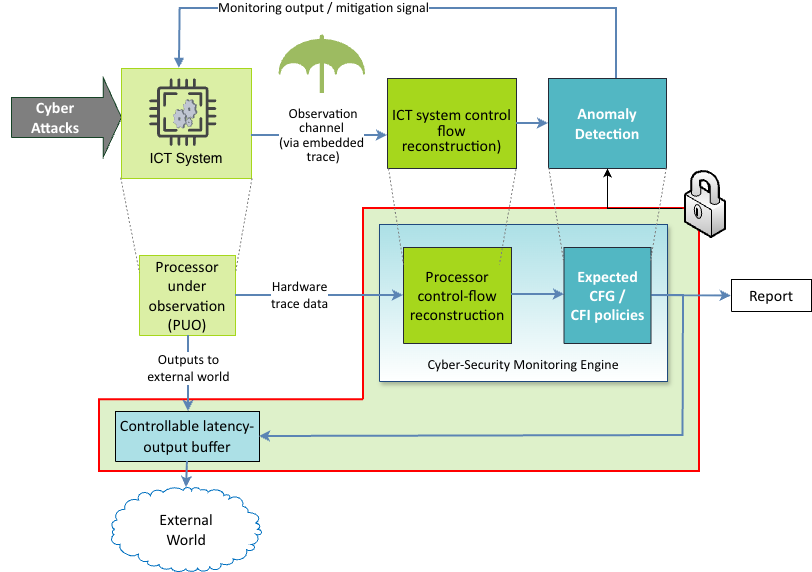}
    \caption{Hardware-supported CFI (Control Flow Integrity) checker (adapted from \cite{weiss2025ep4016343b1}} 
    \label{fig:cfi_checker}
\end{figure*}

In the trace-based approach,
malware has fewer opportunities to suppress or falsify the monitoring evidence
than in software-only approaches as the analysis is
separate from the target system, uses specialized hardware and cannot easily be compromised.
Several hardware-based monitoring architectures along these lines have been proposed in
the literature: \cite{davi2014hardwareassistedcfi} focus on security hardware mechanisms to enable fine-grained CFI checks to protect embedded systems. Kargos \cite{Moon2017} monitors an OS kernel from outside the CPU through the bus
interconnect and the debug interface to detect kernel-level code-injection attacks with
near-zero performance cost. Lee et al. \cite{Lee2017} integrate code-reuse-attack monitoring intellectual property blocks into an ARM-based system-on-chip and feed them from the CoreSight
Program Trace Macrocell, demonstrating that off-CPU CFI monitoring on production trace
interfaces is feasible.
A concrete instantiation of this architecture, which is the basis for the figure and the
anomaly criteria used in the remainder of this paper, is described in \cite{weiss2025ep4016343b1,weiss2025us12462017b2} and shown in
Figure~\ref{fig:cfi_checker}. It comprises three blocks that are jointly responsible for non-intrusive monitoring
and for containing the effects of a detected attack:
\begin{description}

\item[\textnormal{\textit{Processor under observation (PUO):}}]
the system to be protected. It executes the application program code and produces application output data through its regular output interface.
An \textit{observation unit} that is connected to the processing unit non-intrusively monitors the execution and provides \emph{execution information} indicative of the order in which basic blocks and branches are executed. In practice, the observation unit is realized as a trace unit (e.g.\ Arm CoreSight Program Flow Trace \cite{armcoresightpfta} or Intel Processor Trace \cite{intelprocessortrace}) emitting a compressed trace-data stream.

\item[\textnormal{\textit{Cyber-security monitoring engine:}}]
a separate analysis subsystem that is physically separated from and \emph{inaccessible} by the PUO, so that the PUO cannot inspect or alter it even after compromise. It contains a \emph{control-flow reconstruction} unit that decompresses the trace stream and recovers the executed basic-block path, a memory holding a pre-generated model of the expected execution pattern of the uncompromised program, and an \emph{execution-monitoring unit} that compares the reconstructed control flow against this model under one or more explicit anomaly criteria (illegal indirect-branch targets, missing or unexpected source/destination pairs of branches, instruction addresses outside permitted ranges, deviations in the order of execution, exceeded loop/repetition bounds, or call/return mismatches; see Section \ref{sec:hW-monitoring}). When a criterion is violated, it emits a monitoring output signal.

\item[\textnormal{\textit{Controllable latency-output-buffer:}}]
a unit, again inaccessible by the PUO, that sits in the PUO's output path. It temporarily withholds the PUO's output for a bounded latency time span. The output is released only if no anomaly signal has been received during this window; if an anomaly is reported, the output is blocked. The latency time span can be a fixed value (sized to an upper bound of the engine's analysis time) or driven event-based by a buffer-synchronisation signal from the engine.

\end{description}

The latency-output-buffer turns the architecture from a pure \emph{detection} mechanism into a \emph{containment} mechanism: a compromised PUO cannot make its malicious output effective in the outside world before the independent monitor has had a chance to inspect the corresponding execution evidence. For the diagnostic problem studied in this paper, the role of this architecture is more specific: it supplies the attack tree with control-flow evidence that an attacker cannot easily shape from inside the PUO.

Figure~\ref{fig:cfi_checker} shows one concrete containment-oriented instantiation of the hardware-supported CFI architecture from prior work. In the present paper, only the independent trace-based observation and control-flow reconstruction path is essential; the latency-output buffer and mitigation signal illustrate one possible use of the same monitoring evidence for containment.

\section{Walk-Through Example: Authentication Service}
\label{sec:example}

To illustrate the concepts, we consider a small authentication service used by an industrial gateway. The
service receives a login request, checks a password, optionally executes a
maintenance/debug path, checks multi-factor authentication (MFA), and finally
creates a session token.
%
We define the following basic blocks for the example system:

\begin{center}
\begin{tabular}{ll}
Block & Meaning \\
\midrule
$A$ & Receive request \\
$B$ & Parse request \\
$C$ & Check password \\
$D$ & Check whether request is a maintenance/debug request \\
$E$ & Write diagnostic/audit log \\
$F$ & Verify signed maintenance token \\
$G$ & Execute maintenance action \\
$H$ & Check MFA \\
$I$ & Create authenticated session \\
$J$ & Return denial or error \\
$K$ & Decode attacker-controlled payload \\
$L$ & Execute injected payload / patch process state \\
$M$ & Suppress or rewrite software telemetry \\
\end{tabular}
\end{center}

The normal system operation is simple and involves only blocks $A$ to $J$: ordinary users may reach block $I$ only after both
password verification and MFA verification. The maintenance path may write logs
and execute restricted maintenance actions, but only after a valid signed
maintenance token is verified.

\subsection{Nominal Control Flow}

\begin{figure}[htb]
\centering
    \captionsetup{skip=12pt}
\begin{tikzpicture}[
    >=Latex,
    font=\small,
    node distance=4mm and 6mm,
    box/.style={
        draw,
        rounded corners=2mm,
        align=center,
        inner sep=3pt,
        font=\footnotesize
    },
    lab/.style={font=\scriptsize, inner sep=1pt}
]

\node[box] (A) {$A$\\Receive};
\node[box, right=6mm of A] (B) {$B$\\Parse};
\node[box, right=6mm of B] (C) {$C$\\Password};
\node[box, right=13mm of C] (H) {$H$\\MFA};
\node[box, right=13mm of H] (I) {$I$\\Session};

\node[box, above=9mm of C] (D) {$D$\\Debug?};
\node[box, right=14mm of D] (E) {$E$\\Audit log};
\node[box, right=16mm of E] (F) {$F$\\Verify token};
\node[box, right=11mm of F] (G) {$G$\\Maintenance};

\node[box, below=9mm of C] (J) {$J$\\Deny};

\draw[->] (A) -- (B);
\draw[->] (B) -- (C);
\draw[->] (C) -- node[lab, above] {success} (H);
\draw[->] (H) -- node[lab, above] {success} (I);

\draw[->] (C) -- node[lab, left] {failure} (J);

\draw[->] (B.north) |- node[lab, pos=0.25, left] {debug} (D.west);

\draw[->] (D) -- (E);
\draw[->] (E) -- (F);
\draw[->] (F) -- node[lab, above] {valid} (G);

\draw[->] (F) |- node[lab, pos=.2, right] {invalid} (J);

\coordinate (ret) at ($(J.east)+(12mm,0)$);

\draw[->] (I.south) |- (ret) |- (J.east);
\draw[->] (G.south) |- ($(ret)+(12mm,0)$) |- (J.east);

\end{tikzpicture}
\caption{Nominal control-flow graph of the authentication service.}
\label{fig:nominal-cfg}
\end{figure}
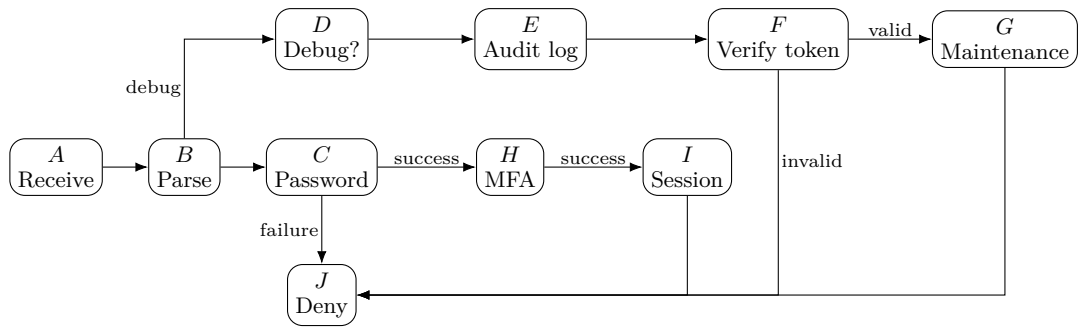

A nominal control-flow graph is shown in
Figure~\ref{fig:nominal-cfg}. Ordinary login follows
\cfg{A,B,C,H,I}. Failed login follows \cfg{A,B,C,J}. Authorized maintenance
follows the path \cfg{A,B,D,E,F,G,J} and does not create a user session.



A control-flow anomaly detector can be defined over observed block sequences.
For example, it may flag an exemption when:
\begin{enumerate}
  \item a session is created without a preceding MFA block;
  \item the maintenance path is entered by an ordinary user;
  \item an indirect branch target is outside the set of known basic blocks; or
  \item a path contains blocks that do not occur in the static control-flow graph.
\end{enumerate}

This detector is intentionally abstract. The example only requires that the
defender obtains some observation 
of the executed control flow and compares
it with a model of expected control flow.

\subsection{Attack Tree for Diagnosis}

\begin{figure}[htb]
\centering
    \captionsetup{skip=12pt}
\begin{tikzpicture}[
  edge from parent/.style={draw, -Latex},
  every node/.style={align=left, font=\footnotesize},
  level 1/.style={sibling distance=44mm, level distance=16mm},
  level 2/.style={sibling distance=22mm, level distance=16mm},
  root/.style={
    draw,
    rounded corners,
    inner sep=3pt,
    text width=56mm
  },
  cat/.style={
    draw,
    rounded corners,
    inner sep=3pt 
  },
  leaf/.style={
    inner sep=1pt,
    text width=22mm
  }
]

\node[root] {Root: Session created without valid MFA}
  child {node[cat] {Benign or low-severity cause}
    child {node[leaf] {MFA exemption misconfigured}}
    child[level distance=28mm] {node[leaf] {Maintenance/debug workflow misuse}}
    child {node[leaf] {Log-only anomaly or tracing bug}}
  }
  child[sibling distance=14mm] {node[cat] {Credential or token abuse}
    child[level distance=28mm] {node[leaf] {Stolen session cookie}}
    child[level distance=28mm] {node[leaf] {Replay of valid service token}}
  }
  child {node[cat] {High-severity intrusion}
    child {node[leaf] {Request parameter tampering}}
    child[level distance=18mm] {node[leaf] {Code injection or ROP}}
    child {node[leaf] {Telemetry suppression or log rewriting}}
  };

\end{tikzpicture}

\caption{A simplified attack tree for interpreting the control-flow anomaly MFA bypass.}
\label{fig:attack-tree}
\end{figure}
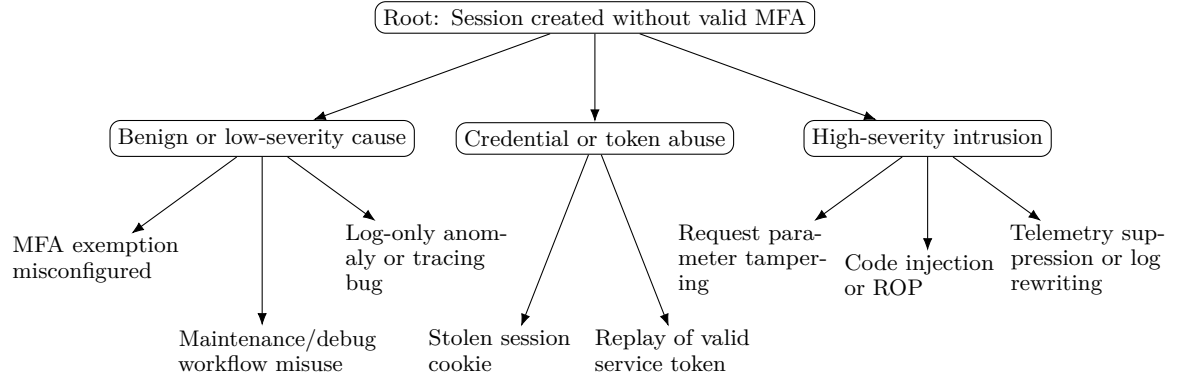

The defender uses an attack tree to interpret observed anomalies.
%
%
Figure~\ref{fig:attack-tree} shows a simplified attack tree for interpreting the anomaly that a session was created without valied MFA. This tree (all internal nodes are disjunctive here) distinguishes between relatively benign
or low-severity explanations, such as a configuration error, and high-severity
intrusions, such as code injection or control-flow hijacking.

The tree can be used as a diagnostic model in the following way: 
observing or not observing anomalies makes its root node true or false, and this can activate or deactivate internal nodes according to the logic of the tree.

For example, the trace \cfg{C,H,I} corresponds to a normal successful login, and no anomaly is detected, so no internal node becomes activated. Similarly, \cfg{D,E,F,G,J} corresponds to the maintenance/debug path being executed normally.  The trace \cfg{C,I} corresponds to a possible MFA bypass, where a session is created without valid MFA, and several internal nodes (such as 'Credential or tokan abuse') can possibly be activated.

Similar to figure~\ref{fig:attack-tree}, attack trees for other anomalies can be defined; however, they are not shown here due to space limitations.  As mentioned in Section \ref{subsec:attack-trees}, several attack trees may be combined straightforwardly into an attack graph with multiple root nodes and shared sub-trees.
In the following, we omit this tree in the presentation and focus only on the feature vector of present or absent anomalies, assuming they correspond to root nodes of the tree.



\subsection{Attack: Control-Flow Hijack Hidden in a Debug Path}

Assume that the attacker exploits a memory-safety bug in the request parser. The exploit is
triggered by a malformed header that is syntactically similar to a legitimate
debug header. The true execution is then $\tau = \cfg{A,B,D,E,K,L,M,I}$.

The semantics of this execution is as follows:
\begin{enumerate}
  \item the request enters the debug-looking path \cfg{D,E} so that the execution
  appears similar to a normal diagnostic workflow.
  \item instead of verifying a signed maintenance token at $F$, a corrupted return
  address or function pointer transfers control to $K$.
  \item $K$ decodes attacker-controlled data embedded in the request.
  \item $L$ patches theh process state so that a session is created with MFA verification set true even though block $H$ was never executed.
  \item $M$ suppresses or rewrites software telemetry, for example by disabling a
  user-space tracing hook or rewriting a local audit buffer.
  \item the process reaches $I$ and creates a valid authenticated session.
\end{enumerate}


%
%
%
%
%
%
%

The actual exploit might use return-oriented programming, an indirect-call
corruption, dynamic code loading, or another mechanism. The example abstracts
away from the exploit mechanism and focuses on the diagnostic consequence: the
true path contains a branch target that is not a valid successor of $E$ in the
nominal control flow.

\subsection{Camouflage: Biasing the Software Observation}

Assume further the attacker's goal is not merely to cause an anomalous path. The attacker wants
the anomaly to be interpreted as a harmless or routine deviation. The camouflage
therefore has two components:
\begin{description}
\item[\textnormal{\textit{Semantic Camouflage.}}]
The malicious transition is placed after the debug/audit blocks \cfg{D,E}. This
makes the beginning of the trace resemble a legitimate maintenance path. From a
coarse observer's perspective, the attack appears to be related to diagnostics,
not to session creation.

\item[\textnormal{\textit{Telemetry Camouflage.}}]
Assume the primary software monitor is implemented in the same operating system
or process environment as the service. For example, it may obtain events from
instrumented logging calls, a user-space tracing library, or an agent running
with ordinary kernel support. The attacker reaches block M before the session is
created and modifies this telemetry channel.
\end{description}


The result is that the software observer records
$\hat{\tau}_S = \cfg{A,B,D,E,F,J}$, followed by a session-create event with missing correlation data.

This observation is inconsistent with nominal control flow, but not maximally suspicious. It suggests that
some maintenance/debug handling occurred, that the request was denied, and that a
later session event lacks proper correlation. Such evidence can be explained by a
benign telemetry problem, a misconfigured debug path, or a confused-deputy bug in
maintenance handling.

\subsection{Wrong Diagnosis from the Attack Tree}

Given \(\hat{\tau}_S\), the attack tree receives the following feature vector of anomalies:

\begin{center}
\begin{tabular}{p{0.56\linewidth}p{0.3\linewidth}}
Anomaly from software observation & Value \\
\midrule
Debug or maintenance path observed & true \\
Signed-token verification block apparently observed & true \\
Unknown branch target observed & false \\
Telemetry discontinuity around malicious block & false or unavailable \\
Session created without a complete login trace & true \\
Evidence of code injection / illegal control transfer & false \\
\end{tabular}
\end{center}

Under these features, the attack tree may rank the most likely diagnosis as:

\begin{quote}
  \textbf{Diagnosis from software observation:} maintenance/debug workflow misuse
  or correlation failure in logging. Severity: medium or low. Root cause:
  misconfiguration, incomplete audit correlation, or an incorrectly exposed
  diagnostic endpoint.
\end{quote}

Because the attack tree reasons only over observed evidence, it reaches a wrong diagnostic conclusion if supplied with this
misleading evidence.
%
%
In particular, the absence of an unknown branch target is a
false negative caused by telemetry camouflage. The attack tree therefore selects
an innocuous branch because the observation channel has been compromised.

\subsection{Observations from Hardware-based Monitoring}\label{sec:hW-monitoring}

Now suppose the platform includes a hardware-based control-flow monitor. The
monitor may be realized using processor trace facilities, a separate monitor
core, a bus-level observer, a trusted execution monitor, or a dedicated hardware
unit that records branch outcomes and indirect branch targets. The essential
property in this example is independence: the monitor is not implemented by the
same software stack that the attacker has compromised.

The hardware monitor records a lower-level trace such as: $\hat{\tau}_H = \cfg{A,B,D,E,K,L,M,I}$.

In practice, the hardware trace may not directly name source-level blocks; it may
record compressed branch decisions, target addresses, exceptions, context
switches, or taken/not-taken outcomes.
In an implementation, unexpected blocks like $K$, $L$ or $M$ may thus initially appear simply as illegal target addresses or unknown basic-block identifiers. The detailed trace decoding and address-to-block mapping are hardware- and toolchain-dependent implementation issues and are therefore not discussed further here.

Assume the defender successfully reconstructs the corresponding
basic-block path by mapping trace packets or branch targets back to the program
image and its expected CFG. The key observation is then the illegal transition:
\[
E \not\rightarrow K \quad \text{in the nominal CFG.}
\]

This is exactly the kind of evidence that the anomaly criteria of the
hardware-based architecture from Section~\ref{sec:approach} are designed to
detect~\cite{weiss2025ep4016343b1}: the observed (source, destination) pair
$(E, K)$ is not contained in the set of permitted branch pairs of
the uncompromised program, and the destination address~$K$ lies outside the
whitelist of permitted indirect-branch targets. In an implementation, the
permitted (source, destination) pairs can be organised as a hash table for
constant-time lookup against each branch packet decoded from the trace stream.

The hardware observation also reveals that H, the MFA block, was not executed:
\[
H \notin \hat{\tau}_H \quad \text{and} \quad I \in \hat{\tau}_H.
\]

This corresponds to a violation of the \emph{order-of-execution} criterion: the
expected program model requires $H \prec I$ on every successful-login path, and
the reconstructed trace does not satisfy this. Note that the same reconstructed trace
could further be checked against a call/return-matching criterion, which
maintains a shadow call stack of return addresses recorded at call instructions
and compares each observed return target against the corresponding entry; in
ROP- or indirect-call-corruption scenarios, this check would already flag the
transfer to~$K$ independently of the (source, destination) whitelist, but such more advanced CFI methods are beyond our scope here.

Finally, it reveals the telemetry manipulation block $M$ before the session
creation:
\[
M \prec I.
\]

Thus, the discrepancy between software and hardware observations is itself
evidence:
\[
\Delta = \hat{\tau}_H \setminus \hat{\tau}_S
       = \{K,L,M,I\ \text{with illegal transition } E\rightarrow K\}.
\]

\subsection{Corrected Attack-Tree Diagnosis}

With \(\hat{\tau}_H\), the feature vector of anomalies changes:

\begin{center}
\begin{tabular}{p{0.46\linewidth}p{0.36\linewidth}}
Feature inferred from hardware observation & Value \\
\midrule
Debug or maintenance path observed & true \\
Signed-token verification block observed & false or bypassed \\
Unknown branch target observed & true \\
Session created without MFA & true \\
Telemetry manipulation before session creation & true \\
Software and hardware observations disagree & true \\
\end{tabular}
\end{center}

The attack-tree diagnosis now changes:

\begin{quote}
  \textbf{Diagnosis from hardware observation:} high-severity intrusion involving
  illegal control transfer from the debug/audit path, followed by payload
  execution and telemetry suppression. Likely root cause: memory-safety bug or
  indirect-call corruption in the debug handler, used to bypass MFA and create a
  session.
\end{quote}

The same attack tree therefore produces two different diagnoses depending on the
quality of the control-flow evidence. 
We summarize the example in the table below.
\begin{center}
\small
\begin{tabular}{p{0.22\linewidth}p{0.30\linewidth}p{0.36\linewidth}}
Scenario & Observation & Attack-tree result \\
\midrule
Nominal execution & \cfg{A,B,C,H,I} & No attack \\
True attack & \cfg{A,B,D,E,K,L,M,I} & Code injection / control-flow hijack with MFA bypass \\
Camouflaged software observation & \cfg{A,B,D,E,F,J} plus poorly correlated session event & Maintenance/debug misuse or logging correlation failure \\
Hardware observation & \cfg{A,B,D,E,K,L,M,I} with illegal transition \cfg{E,K} and missing $H$ & High-severity intrusion: illegal branch target, payload execution, telemetry suppression, MFA bypass \\
\end{tabular}
\end{center}

\section{Discussion and Ongoing Work}

As the example illustrated, hardware-based monitoring does not merely improve anomaly detection; it improves
  the diagnostic value of attack-tree analysis by improving the fidelity of the
  control-flow evidence supplied to the tree. When software telemetry is
  camouflaged, the attack tree may identify a harmless root cause. When an
  independent hardware monitor supplies the actual path, the same tree can select
  a more faithful high-severity attack class.

Beyond diagnosis, the architecture sketched in Section~\ref{sec:approach}
also enables a form of \emph{containment}: by interposing a latency-output-buffer
between the PUO's output interface and the outside world, the system can withhold
output until the independent monitor has had the opportunity to detect an
anomaly~\cite{weiss2025ep4016343b1,weiss2025us12462017b2}. A correct attack-tree
diagnosis based on hardware evidence can in this case trigger a meaningful
mitigation (e.g.\ blocking the suspect output, raising an alert, or transitioning
the system into a safe state) before the consequences of the intrusion become
externally visible (see also Figure~\ref{fig:cfi_checker}).

While our work illustrates the beneficial interplay of control-flow anomaly detection, attack trees, and hardware-based monitoring, it is still preliminary as a number of challenges remain. Especially, a richer theory of attack types, attack camouflage, and how this affects monitorability and diagnosability in the context of security needs to be developed. In the area of runtime verification, there is a body of work on the concept of partial observability and imperfect/incomplete traces
that is closely related to this problem \cite{Cimatti2019,Ferrando2022,taleb2023uncertainty}.
Incomplete or partially available traces have also been studied for stream
runtime verification, where gaps in the input stream are represented
abstractly as sets of possible concrete traces \cite{Leucker2019}.

It should also be pointed out that we do not claim that the approach of hardware-based trace monitoring is intrinsically perfect, since it
can itself suffer from configuration errors, trace loss, aliasing, incomplete coverage,
or physical attacks:
\begin{itemize}
  \item Hardware traces may be incomplete if the trace buffer overflows, if tracing
  is incorrectly configured, or if relevant execution occurs outside the monitored
  context.
  \item Mapping low-level branch traces to source-level basic blocks can be
  difficult in the presence of dynamic loading, JIT compilation, self-modifying
  code, interrupts, or aggressive compiler optimization.
  evidence, but attack-tree diagnosis still requires a model of possible attacks
  and root causes.
  \item A sufficiently powerful adversary may attack the hardware monitor,
  firmware, trace configuration, or physical platform. The security claim depends
  on the monitor being outside the adversary's effective control.
  \end{itemize}
Also, processing hardware-level trace information at run-time comes at significant computational and bandwith cost, and it is therefore unrealistic (and unnecessary) to perform hardware monitoring continuously along software monitoring. Instead, it is desirable to activate this lower-level monitoring only when really necessary, such as in the presence of suspicious activities.
 For instance, in a scenario with a server that has several CPU cores, as soon as software monitoring flags a CPU core as suspicious, hardware monitoring could focus on  this part of the system.
We are thus working on a tighter diagnostic integration, where software-level observation provides a focus by indicating suspicious activities, while hardware-level monitoring is used to check these candidate incidents in more detail.

  \section{Conclusion}
We presented an approach towards run-time cybersecurity that guards software execution against flows of control not intended by the original program. 
This is achieved by monitoring program flow, and mapping observed deviations against a structured representation of hypothesized attacks (attack trees).
A key complication (and novel aspect in diagnosis) is that attackers might actively and adversarially influence observations in order to conceal their actions; we seek to address this camouflage in our architectural framework through incorporating hardware-level trace information, an observation channel
implemented outside the compromised software stack that can reveal discrepancies that software-only measures may miss.
Though several challenges like a more rigorous formalization and analysis still remain, the approach represents a step towards model-based diagnosis concepts for cybersecurity.



\bibliography{references}

\end{document}